# The quantum mechanics based on a general kinetic energy

Yuchuan Wei[*]

International Center of Quantum Mechanics, Three Gorges University, China, 443002

Department of Radiation Oncology, Wake Forest University, NC, 27157

*Corresponding author: yuchuanwei@gmail.com

Abstract

In this paper, we introduce the Schrödinger equation with a general kinetic energy operator. The conservation law is proved and the probability continuity equation is deducted in a general sense. Examples with a Hermitian kinetic energy operator include the standard Schrödinger equation, the relativistic Schrödinger equation, the fractional Schrödinger equation, the Dirac equation, and the deformed Schrödinger equation. We reveal that the Klein-Gordon equation has a hidden non-Hermitian kinetic energy operator. The probability continuity equation with sources indicates that there exists a different way of probability transportation, which is probability teleportation. An average formula is deducted from the relativistic Schrödinger equation, the Dirac equation, and the K-G equation.

## I. Introduction

Kinetic energy originally appeared in Newton's classical mechanics, and was later revised by Einstein in special relativity. In 1926 [1], Schrödinger viewed the classical kinetic energy as an operator, and discovered that the energy levels of the Hydrogen atom were eigenvalues of the Hamiltonian operator. In 1929 [2], Dirac expressed the relativistic kinetic energy as a 4-dimesional linear operator, and obtained the energy levels with fine structure of the H atom. Recently [3,4], based on the relativistic kinetic energy operator itself, we obtained a new energy formula with an $\alpha^5$ term. Together with the effect of spin-orbit coupling, the new energy formula can explain the Lamb shift half-quantitatively.

Additionally, the fractional kinetic energy appears in the fractional quantum mechanics [5-8], and the kinetic energy with a position-dependent mass appears in the deformed quantum mechanics [9,10].

In order to build a unified framework for different equations in quantum mechanics, we will introduce the Schrödinger equation with a general kinetic energy operator. In a general sense, the conservation law is proved and the probability continuity equation is derived. The standard Schrödinger equation, the fractional Schrödinger equation, the relativistic Schrödinger equation, the reformed Schrödinger equations, the Dirac equation, and the K-G equation are considered as concrete examples. A new way of probability transportation, probability teleportation, is proposed.

## II. Various kinetic energy operators in quantum mechanics

We first recall the various kinetic energy operators used in quantum theory through the history [1,2], and explain the fractional kinetic energy operator in the fractional quantum mechanics [5-6] and the kinetic energy operator with a position-dependent mass in the deformed quantum mechanics[9,10].

### 1. The formulas for the energy levels of the hydrogen atom and the related kinetic energy operators

Historically, the explanation of the hydrogen spectrum has been the target of quantum theory and a better formula for the hydrogen energy levels has been the purpose of theoretical physicists [1,2]. We will see that different kinetic energy operators lead to different formulas for the energy levels of the H atom.

In classical mechanics, the total energy H (the Hamiltonian) of the hydrogen atom was the summation of the classical kinetic energy T and the Coulomb potential energy V of the electron

$$H = T + V, \quad T = \frac{\mathbf{p}^2}{2m}, \quad V = -\frac{e^2}{r}. \tag{1}$$

As usual, *m* and *e* are the mass and charge of the electron, and **p** is the momentum of the electron.

Under a new assumption that the electron ran only along circular orbits with angular moment $n\hbar$, in 1913 Bohr obtained the first formula of the energy levels

$$E = -mc^2\alpha^2 \frac{1}{2n^2}, \quad n=1,2,3\cdots . \tag{2}$$

As usual, c is the speed of light, and α is the fine structure constant.

In order to further explain the fine structure of the hydrogen spectrum, Sommerfeld pointed out that the relativistic kinetic energy should be used in the Hamiltonian

$$H = T + V, \quad T = \sqrt{\mathbf{p}^2 c^2 + m^2 c^4}, \quad V = -\frac{e^2}{r} \tag{3}$$

and that elliptic orbits should be allowed as well. The revised energy level formula became

$$E = mc^2 \left( 1 + \frac{\alpha^2}{\left(n_r + \sqrt{k^2 - \alpha^2}\right)^2} \right)^{-1/2} \tag{4}$$

with $n_r = 0,1,2,\cdots$, $k = 1,2,3\cdots$. This formula matched the fine structure of the hydrogen spectrum very well.

In 1926, Schrödinger discovered that the Bohr energy levels can be obtained by solving the eigenequation of the Hamiltonian operator

$$H\psi = E\psi, \quad H = -\frac{\hbar^2}{2m}\nabla^2 - \frac{e^2}{r}. \tag{5}$$

This was the beginning of quantum mechanics. It was a natural wish that we could get the energy level formula for the fine structure from the eigenequation

$$H\psi = E\psi, \quad H = \sqrt{\mathbf{p}^2 c^2 + m^2 c^4} - \frac{e^2}{r}. \tag{6}$$

However, it was not easy to solve this equation, so usually the Hamiltonian was approximated as

$$H = m^2 c^4 + \frac{\mathbf{p}^2}{2m} - \frac{1}{8}\frac{\mathbf{p}^4}{m^3 c^2} - \frac{e^2}{r} + \cdots . \tag{7}$$

Based on the perturbation theory, the energy levels of the first order approximation were

$$E = mc^2 \left(1 - \frac{1}{2n^2}\alpha^2 - \frac{1}{2n^4}(\frac{n}{l+1/2} - \frac{3}{4})\alpha^4\right), \qquad (8)$$

which unfortunately did not match the fine structure of the Hydrogen spectrum. Here $n = 1, 2, \cdots$, $l = 0, 1, 2, \cdots$.

Considering the relativistic covariance, the Klein-Gordon equation was proposed in 1926 as well. From the K-G equation [1,2]

$$(E - \frac{e^2}{r})^2 \psi = (\mathbf{p}^2 c^2 + m^2 c^4)\psi \qquad (9)$$

we got the energy level formula

$$E = mc^2 \left[1 + \frac{\alpha^2}{\left(n_r + \frac{1}{2} + \sqrt{(l+1/2)^2 - \alpha^2}\right)^2}\right]^{-1/2}$$

$$= mc^2 \left(1 - \frac{1}{2n^2}\alpha^2 - \frac{1}{2n^4}(\frac{n}{l+1/2} - \frac{3}{4})\alpha^4 + \cdots\right) \qquad (10)$$

This formula had the same $\alpha^4$ term as the result of the perturbation method (8), which did not match the fine structure of the Hydrogen spectrum either. In section V, we will reveal the kinetic energy operator hidden in the K-G equation.

In 1929 [2], using a 4-dimensional linear kinetic energy operator, Dirac proposed his equation

$$H\psi = E\psi, \quad H = c\vec{\alpha} \cdot \mathbf{p} + \beta mc^2 - e^2/r, \qquad (11)$$

where $\vec{\alpha}, \beta$ are 4×4 matrices.

The new energy formula was the same as Sommerfeld's formula (4), which matches the fine structure of the hydrogen spectrum very well.

In 1947 [1,2], Lamb discovered that the states $2S_{1/2}$ and $2P_{1/2}$ of the H atom, which should have the same energy according to Dirac's equation, actually had an about 1000Mhz microwave radiation between them. Theoretical physicists believed that the Lamb shift was beyond the scope of quantum mechanics and they developed quantum electrodynamics to explain it.

Recently we revisited the popular perturbation method to deal with the kinetic relativistic correction and found some basic problems [3,4]. If we keep the $\mathbf{p}^4$ term only, i.e.,

$$H = m^2c^4 + \frac{\mathbf{p}^2}{2m} - \frac{1}{8}\frac{\mathbf{p}^4}{m^3c^2} - \frac{e^2}{r} \tag{12}$$

its eigenvalues do not exist, since the kinetic energy becomes negative as the momentum magnitude increases very large. On the other hand, if we include $\mathbf{p}^6$ term as well, some of the matrix elements of the $\mathbf{p}^6$ term are infinity. Thus, it was impossible to find the relativistic correction of the energy levels of the H atom by using the power series of the relativistic kinetic energy and the relativistic correction of the kinetic energy of the H atom has never been found in a reasonable way.

Therefore, we developed a perturbation method based on the square root operator directly, rather than its power expansion, and calculated the energy levels carefully. The new energy levels contain a valuable $\alpha^5$ term. Together with the L-S coupling effect, the new energy formula

$$E = mc^2\left[1 - \frac{1}{2n^2}\alpha^2 - \frac{1}{2n^4}\left(\frac{n}{j+\frac{1}{2}} - \frac{3}{4}\right)\alpha^4 + \frac{64}{15\pi}\frac{1}{n^3}\alpha^5\delta_{0l} + O(\alpha^6)\right] \tag{13}$$

can half-quantitatively explain the Lamb shift [3]. Here $n = 1, 2, 3\cdots$, $j + 1/2 = 1, 2, 3\cdots$, $l = 0, 1, 2\cdots$, $\delta_{00} = 1$, $\delta_{01} = \delta_{02} = \cdots = 0$.

In one word, the history of quantum theory shows us that various kinetic energy operators lead to different formulas for energy levels of the H atom, and the relativistic kinetic energy operator plays a special role.

## 2. Fractional kinetic energy

In the fractional quantum mechanics [5-6], the fractional kinetic energy operator is expressed as

$$T_\alpha = D_\alpha |\mathbf{p}|^\alpha = D_\alpha\left(-\hbar^2\nabla^2\right)^{\alpha/2}, \tag{14}$$

where the fractional parameter $1 < \alpha \leq 2$, and $D_\alpha$ is a coefficient dependent on $\alpha$. Please be reminded of the different meanings of the three symbols, $\alpha$ (the fractional parameter), α (the fine structure constant), and $\vec{\alpha}$ (4×4 matrices).

In the case $\alpha = 2$, taking $D_2 = 1/(2m)$, the fractional kinetic energy is the classical kinetic energy

$$T_2 = \frac{\mathbf{p}^2}{2m}. \tag{15}$$

### 3. The kinetic energy with a position-dependent mass.

In the deformed Schrödinger equation [9,10], the mass $m(\mathbf{r})$ is a position dependent function and the kinetic energy operator is

$$T = \mathbf{p} \cdot \frac{1}{2m(\mathbf{r})} \mathbf{p}. \tag{16}$$

Now that there are various types of quantum mechanics based on different kinetic energy operators, it is necessary to introduce the quantum mechanics based on a general kinetic energy.

## III. Quantum mechanics based on a general kinetic energy

In this section we propose the Schrödinger equation with a general kinetic energy, prove the conservation law, and derive the probability continuity equations in a general sense.

### 1. The Schrödinger equation with a general kinetic energy operator

All the square integrable functions defined on the three-dimensional Euclidian space $R^3$ form a Hilbert space with an inner product

$$<\varphi, \chi> = \int_{R^3} \varphi^*(\mathbf{r}) \chi(\mathbf{r}) d^3\mathbf{r}, \tag{17}$$

where $\varphi, \chi$ are any two square integrable functions. $\mathbf{r} \in R^3$ is a three-dimensional vector. In quantum mechanics, square integrable functions are called wavefunctions[1].

We call

$$i\hbar \frac{\partial}{\partial t} \Psi(\mathbf{r}, t) = H\Psi(\mathbf{r}, t) \tag{18}$$

the general Schrödinger equation, where $\Psi(\mathbf{r}, t)$ is a time-dependent wavefunction.

Here the Hamiltonian operator

$$H = T + V = T(\mathbf{r}, \mathbf{P}, t) + V(\mathbf{r}, t) \tag{19}$$

is a summation of a general kinetic energy operator T and a potential energy operator V.

As a real-valued function, the potential energy is always a Hermitian operator, $V^+ = V$. Usually the kinetic energy operator is Hermitian, i.e., $T^+ = T$, so that the Hamiltonian operator H is Hermitian as well. The quantum system with a Hermitian Hamiltonian is called a Hermitian system. In a Hermitian system, the eigenvalues of the total energy are real and the wavefunctions related to different energies are orthogonal.

Occasionally, the kinetic energy operator is non-Hermitian, e.g., we will reveal that the Klein-Gordon equation has a hidden non-Hermitian Hamiltonian operator. The quantum system with a non-Hermitian Hamiltonian is called a non-Hermitian system.

**2. The conservation law**

Suppose that A is a time-independent operator and that $\psi(\mathbf{r})$ is a square integrable function, we call the integral

$$\bar{A} =< A >=< \psi, A\psi >= \int_{R^3} \psi^*(\mathbf{r}) A\psi(\mathbf{r}) d^3\mathbf{r} \tag{20}$$

the value of the operator A on the wavefunction $\psi(\mathbf{r})$. If A is a Hermitian operator and $\psi(\mathbf{r})$ is normal, $\bar{A}$ is the well-known average. If $\psi(\mathbf{r},t)$ is a solution of the general Schrödinger equation, then

$$\bar{A}(t) = \int_{R^3} \psi^*(\mathbf{r},t) A\psi(\mathbf{r},t) d^3\mathbf{r} \tag{21}$$

is a function of time $t$.

**Theorem 1**. The quantity $\bar{A}$ is conserved when

$$H^+ A - AH = 0. \tag{22}$$

**Proof.**

From (18), we know that the derivative of the quantity $\bar{A}$ is

$$\begin{aligned}\frac{d}{dt}\bar{A}(t) &= \frac{d}{dt}\int_{R^3} \psi^*(\mathbf{r},t) A\psi(\mathbf{r},t) d^3\mathbf{r} \\ &= \int_{R^3} \frac{d}{dt}\psi^*(\mathbf{r},t) A\psi(\mathbf{r},t) d^3\mathbf{r} + \int_{R^3} \psi^*(\mathbf{r},t) A\frac{d}{dt}\psi(\mathbf{r},t) d^3\mathbf{r} \\ &= \int_{R^3} -\frac{1}{i\hbar}(H\psi)^*(\mathbf{r},t) A\psi(\mathbf{r},t) d^3\mathbf{r} + \int_{R^3} \psi^*(\mathbf{r},t) A\frac{1}{i\hbar} H\psi(\mathbf{r},t) d^3\mathbf{r} \\ &= \int_{R^3} -\frac{1}{i\hbar}\psi^*(\mathbf{r},t) H^+ A\psi(\mathbf{r},t) d^3\mathbf{r} + \int_{R^3} \psi^*(\mathbf{r},t) A\frac{1}{i\hbar} H\psi(\mathbf{r},t) d^3\mathbf{r} \\ &= \frac{i}{\hbar}\int_{R^3} \psi^*(\mathbf{r},t)(H^+ A - AH)\psi(\mathbf{r},t) d^3\mathbf{r}.\end{aligned} \tag{23}$$

Obviously, we have

$$\frac{d}{dt}\bar{A} = 0 \text{ when } H^+ A - AH = 0 \tag{24}$$

This completes the proof.

**Theorem 2**. For a Hermitian Hamiltonian, $H^+ = H$, the quantity $\bar{A}$ is conserved if its operator and the Hamiltonian commutes

$$[H, A] = 0. \tag{25}$$

**Proof.** This is a straightforward result of the theorem 1.

**Theorem 3.** The integral $\int_{R^3} \psi^*(\mathbf{r},t)\psi(\mathbf{r},t)d^3\mathbf{r}$ is conserved if $H^+ = H$.

**Proof.** The identity operator is defined as

$$I\chi = \chi. \tag{26}$$

Taking A=I in Theorem 1, we get this theorem at once. This completes the proof.

According to this theorem, in a Hermitian system, the probability is conserved. If at t=0, the wavefunction is normalized

$$\int_{R^3} \psi^*(\mathbf{r},0)\psi(\mathbf{r},0)d^3\mathbf{r} = 1, \tag{27}$$

then the wavefunction is normalized at any time $t$, i.e.

$$\int_{R^3} \psi^*(\mathbf{r},t)\psi(\mathbf{r},t)d^3\mathbf{r} = 1. \tag{28}$$

Reversely, in a non-Hermitian system, it is useless to normalize a wavefunction, since the normalized function will become non-normalized in the next moment.

## 3. The probability continuity equations with and without sources

From the general Schrödinger equation (18) we have

$$i\hbar \psi^* \frac{\partial}{\partial t}\psi = \psi^* T\psi + \psi^* V\psi,$$
$$i\hbar \psi \frac{\partial}{\partial t}\psi^* = -\psi(T\psi)^* - \psi V\psi^*. \tag{29}$$

Adding the above two equations, we have

$$i\hbar \frac{\partial}{\partial t}\psi^*\psi = \psi^*(T\psi) - \psi(T\psi)^*, \tag{30}$$

$$\frac{\partial}{\partial t}\psi^*\psi + \frac{i}{\hbar}\left(\psi^*(T\psi) - \psi(T\psi)^*\right) = 0. \tag{31}$$

The probability continuity equation can be written as

$$\frac{\partial}{\partial t}\rho + \nabla \cdot \mathbf{j} = 0 \tag{32}$$

where the probability density and the current density is defined as

$$\begin{aligned}\rho &= \psi^*\psi \\ \mathbf{j} &= \frac{i}{\hbar}\nabla^{-2}\nabla\left(\psi^*(T\psi) - \psi(T\psi)^*\right).\end{aligned} \tag{33}$$

However, the form of the probability continuity equation is not unique. Generally we can write the probability continuity equation as

$$\frac{\partial}{\partial t}\rho + \nabla \cdot \mathbf{j} = I \tag{34}$$

where the current density and the source term can be defined in several ways

$$\begin{aligned}\mathbf{j}_1 &= \frac{i}{\hbar}\left(\psi^*\nabla^{-2}\nabla(T\psi) - \psi\nabla^{-2}\nabla(T\psi)^*\right) \\ I_1 &= \frac{i}{\hbar}\left(\nabla\psi^*\nabla^{-2}\nabla(T\psi) - \nabla\psi\nabla^{-2}\nabla(T\psi)^*\right).\end{aligned} \tag{35}$$

$$\begin{aligned}\mathbf{j}_2 &= \frac{i}{\hbar}\left((\nabla^{-2}\nabla\psi^*)(T\psi) - (\nabla^{-2}\nabla\psi)(T\psi)^*\right) \\ I_2 &= \frac{i}{\hbar}\left((\nabla^{-2}\nabla\psi^*)(\nabla T\psi) - (\nabla^{-2}\nabla\psi)\nabla T\psi^*\right),\end{aligned} \tag{36}$$

$$\begin{aligned}\mathbf{j}_3 &= 0 \\ I_3 &= -\frac{i}{\hbar}\left(\psi^*(T\psi) - \psi(T\psi)^*\right).\end{aligned} \tag{37}$$

If $I(\mathbf{r},t) > 0$, there is a source at position $\mathbf{r}$ and time $t$, which generates the probability; when $I(\mathbf{r},t) < 0$, there is a sink at position $\mathbf{r}$ and time $t$, which destroys the probability.

Here are the different pictures of the probability transportation in the different systems.

In a non-Hermitian system, the total probability is not conserved: the continuity equation without sources (32) tells us that the probability can generated or destroyed in infinitely far places, but they are conserved in any finitely far place; the continuity equation with sources (34) tell us that the probability can be generated or destroyed everywhere. It is lucky that we seldom encounter the non-Hermitian systems.

In a Hermitian system, the total probability is conserved. (1) The continuity equation without sources (32) tells us that the probability cannot be generated or destroyed anywhere, and it moves from one place to another place. This is the popular picture in our mind on the probability transportation, and the scattering experiments are based on this understanding. (2) However, the continuity equation with sources (34) tells us that the probability can disappear in one place and simultaneously appear in other places, but the total probability does not change. In other words, some probabilities can be teleported from one place to another. Furthermore, if the particle has mass and charge, probability teleportation will imply mass teleportation and charge teleportation. This is a new picture of the probability transportation.

We must further study the different effects caused by different continuity equations and design suitable experiments to decide which picture on the probability transportation is correct.

### IV. Examples of the Hermitian kinetic energy operator

In this section we study the concrete examples of the Hermitian kinetics energy operator, including the standard Schrödinger equation, the relativistic Schrödinger equation, the fractional Schrödinger equation, the Schrödinger equation with a position dependent mass, and the Dirac equation.

### 1. The standard Schrödinger equation

In quantum mechanics [1], the standard Schrödinger equation is

$$i\hbar \frac{\partial}{\partial t} \Psi(\mathbf{r},t) = H\Psi(\mathbf{r},t), \tag{38}$$

where the Hamiltonian operator is

$$H = T + V = -\frac{\hbar^2}{2m}\nabla^2 + V(\mathbf{r},t). \tag{39}$$

The probability continuity equation without source is

$$\frac{\partial}{\partial t}\rho + \nabla \cdot \mathbf{j} = 0 \tag{40}$$

with

$$\begin{aligned}\rho &= \psi^*\psi \\ \mathbf{j} &= -\frac{i\hbar}{2m}\left(\psi^*\nabla\psi - \psi\nabla\psi^*\right).\end{aligned} \tag{41}$$

This well-known result is consistent to the definition (33) as well as the definition (35).

When a spinless particle with mass *m* and charge *q* moves in an electromagnetic field, its nonrelativistic Hamiltonian [1] is

$$H = T + V = \frac{1}{2m}(\mathbf{p} - \frac{q}{c}\mathbf{A})^2 + q\Phi \qquad (42)$$

where $\mathbf{A} = \mathbf{A}(\mathbf{r},t)$ and $\Phi = \Phi(\mathbf{r},t)$ are the electromagnetic potential.

The Pauli's equation is written as

$$i\hbar \frac{\partial}{\partial t}\Psi(\mathbf{r},t) = \frac{1}{2m}(-i\hbar\nabla - q\mathbf{A}/c)^2 \Psi(\mathbf{r},t) + q\Phi\Psi(\mathbf{r},t). \qquad (43)$$

The probability continuity equation without source remains Eq. (40) with a new current density

$$\mathbf{j} = \frac{1}{2m}\psi^*(\hat{\mathbf{p}} - q\mathbf{A}/c)\psi + cc, \qquad (44)$$

where letters cc is the complex conjunction of the term in front of it.

Here is the deduction from the definition (33)

$$\begin{aligned}
\mathbf{j} &= \frac{i}{\hbar}\nabla^{-2}\nabla\left(\psi^*(T\psi) - \psi(T\psi)^*\right) \\
&= \frac{i}{2m\hbar}\nabla^{-2}\nabla\left(\psi^*(\mathbf{p} - q\mathbf{A}/c)^2\psi - \psi\left((\mathbf{p} - q\mathbf{A}/c)^2\psi\right)^*\right) \\
&= \frac{i}{2m\hbar}\nabla^{-2}\nabla\left(\psi^*(\mathbf{p}^2 - 2q\mathbf{A}\cdot\mathbf{p}/c + q^2\mathbf{A}^2/c^2)\psi - \psi\left((\mathbf{p}^2 - 2q\mathbf{A}\cdot\mathbf{p}/c + q^2\mathbf{A}^2/c^2)\psi\right)^*\right) \\
&= \frac{i}{2m\hbar}\nabla^{-2}\nabla\left(\psi^*(\mathbf{p}^2 - 2q\mathbf{A}\cdot\mathbf{p}/c)\psi - \psi(\mathbf{p}^2 + 2q\mathbf{A}\cdot\mathbf{p}/c)\psi^*\right) \\
&= \frac{i}{2m\hbar}\nabla^{-2}\nabla\left(\psi^*\mathbf{p}^2\psi - \psi\mathbf{p}^2\psi^*\right) - \frac{iq}{m\hbar c}\nabla^{-2}\nabla\left(\psi^*\mathbf{A}\cdot\mathbf{p}\psi + \psi\mathbf{A}\cdot\mathbf{p}\psi^*\right) \\
&= \frac{-i\hbar}{2m}\nabla^{-2}\nabla\left(\psi^*\nabla^2\psi - \psi\nabla^2\psi^*\right) - \frac{q}{mc}\nabla^{-2}\nabla\left(\psi^*\mathbf{A}\cdot\nabla\psi + \psi\mathbf{A}\cdot\nabla\psi^*\right) \\
&= \frac{-i\hbar}{2m}\nabla^{-2}\nabla\cdot\nabla\left(\psi^*\nabla\psi - \psi\nabla\psi^*\right) - \frac{q}{mc}\nabla^{-2}\nabla\cdot\nabla\left(\psi^*\mathbf{A}\cdot\psi\right) \\
&= \frac{-i\hbar}{2m}\left(\psi^*\nabla\psi - \psi\nabla\psi^*\right) - \frac{q}{mc}\left(\psi^*\mathbf{A}\psi\right) \\
&= \frac{1}{2m}\left(\psi^*\mathbf{p}\psi - \psi\mathbf{p}\psi^*\right) - \frac{q}{mc}\left(\psi^*\mathbf{A}\psi\right) \\
&= \frac{1}{2m}\psi^*(\mathbf{p} - q\mathbf{A}/c)\psi + cc.
\end{aligned} \qquad (45)$$

Here we have used Coulomb gauge of the electromagnetic field

$$\nabla \cdot \mathbf{A} = 0,$$
$$\mathbf{p} \cdot \mathbf{A}\psi = \mathbf{A} \cdot \mathbf{p}\psi. \tag{46}$$

On the other hand, according to the definition (35), the continuity equation can be written as

$$\frac{\partial}{\partial t}\rho + \nabla \cdot \mathbf{j}_1 = I_1 \tag{47}$$

where

$$\mathbf{j}_1 = \frac{1}{2m}\left(\psi^*(\hat{\mathbf{p}} - 2q\mathbf{A}/c)\psi + cc\right), \tag{48}$$

$$I_1 = -\frac{q}{mc}\nabla \cdot \left(\psi^*\mathbf{A}\psi\right). \tag{49}$$

As we said above, the difference between the two continuity equations should be studied carefully in theory and experiment.

## 2. Relativistic Schrödinger equation.

1) Without magnetic field

According to special relativity, the kinetic energy is

$$T = \sqrt{\mathbf{p}^2 c^2 + m^2 c^4}. \tag{50}$$

For a particle moving in a potential field V, the relativistic Hamiltonian function is [11,12]

$$H = \sqrt{\mathbf{p}^2 c^2 + m^2 c^4} + V(\mathbf{r}). \tag{51}$$

The relativistic Schrödinger equation is

$$i\hbar\frac{\partial}{\partial t}\psi(\mathbf{r},t) = H\psi(\mathbf{r},t). \tag{52}$$

We can easily prove that if the potential is even, $V(-\mathbf{r},t) = V(\mathbf{r},t)$, then the parity is conserved and that if the potential is central $V(\mathbf{r},t) = V(r,t)$, then the angular momentum is conserved.

As usual, the parity operator P and the angular momentum **L** are defined as

$$P\psi(\mathbf{r}) = \psi(-\mathbf{r}),$$
$$\mathbf{L} = \mathbf{r} \times \mathbf{p}. \tag{53}$$

**Formula 1.** If an eigenvalue and an eigenfunction of the Hamiltonian operator (51) are E and $\psi(\mathbf{r})$, i.e.

$$H\psi(\mathbf{r}) = E\psi(\mathbf{r}), \tag{54}$$

then we have

$$E = <V> + \sqrt{<\mathbf{p}^2 c^2> + m^2 c^4 + <V>^2 - <V^2>}. \tag{55}$$

**Proof.**

From (51), we have

$$H - V = \sqrt{\mathbf{p}^2 c^2 + m^2 c^4} \tag{56}$$

$$(H - V)^2 = \mathbf{p}^2 c^2 + m^2 c^4 \tag{57}$$

$$H^2 - HV - VH + V^2 = \mathbf{p}^2 c^2 + m^2 c^4. \tag{58}$$

Taking the integral $\int_{R^3} \psi^*(\mathbf{r}) \cdots \psi(\mathbf{r}) d^3\mathbf{r}$ of the above equation, we have

$$\int_{R^3} \psi^*(\mathbf{r})(H^2 - HV - VH + V^2)\psi(\mathbf{r}) d^3\mathbf{r} = \int_{R^3} \psi^*(\mathbf{r})(\mathbf{p}^2 c^2 + m^2 c^4)\psi(\mathbf{r}) d^3\mathbf{r} \tag{59}$$

Using the Dirac's notations, we have

$$\begin{aligned}
&<H^2> - <HV> - <VH> + <V^2> = <\mathbf{p}^2 c^2> + m^2 c^4 \\
&E^2 - 2E<V> + <V^2> = <\mathbf{p}^2 c^2> + m^2 c^4 \\
&(E - <V>)^2 = <\mathbf{p}^2 c^2> + <V>^2 - <V^2> + m^2 c^4
\end{aligned} \tag{60}$$

$$E = <V> + \sqrt{<\mathbf{p}^2 c^2> + <V>^2 - <V^2> + m^2 c^4}. \tag{61}$$

We keep the positive solution only in the above equation according to the meaning in physics. This completes the proof.

Since $E = <V> + <T>$, we have the average of the kinetic energy

$$<T> = \sqrt{<\mathbf{p}^2 c^2> + <V>^2 - <V^2> + m^2 c^4}. \tag{62}$$

This relation is similar to the Virial theorem in the standard quantum mechanics [1].

**2) With magnetic fields**

More generally, if a relativistic particle with mass $m$ and charge $q$ moves in an electromagnetic field, its Hamiltonian is

$$H(\mathbf{r},\mathbf{p},t) = \sqrt{(\mathbf{p}c - q\mathbf{A}(\mathbf{r},t))^2 + m^2c^4} + q\Phi(\mathbf{r},t) \qquad (63)$$
$$= T(\mathbf{r},\mathbf{p},t) + V(\mathbf{r},t).$$

One may define the kinetic energy operator $T(\mathbf{r},\mathbf{p},t)$ as follows. At any given time $t$, suppose that the eigenequation of the operator $(\mathbf{p}c - q\mathbf{A}(\mathbf{r},t))^2$

$$(\mathbf{p}c - q\mathbf{A}(\mathbf{r},t))^2 \psi_{\lambda(t)}(\mathbf{r}) = \lambda(t)\psi_{\lambda(t)}(\mathbf{r}) \qquad (64)$$

has been solved, i.e., the eigenfunctions $\psi_{\lambda(t)}(\mathbf{r})$ and eigenvalues $\lambda(t)$ are known. Since the operator $\mathbf{p}c - q\mathbf{A}(\mathbf{r},t)$ is Hermitian, $(\mathbf{p}c - q\mathbf{A}(\mathbf{r},t))^2$ is a nonnegative-definite operator, and $\lambda(t)$ is a nonnegative number. As a simple example, the eigenfunctions and eigenvalues of (64) for a static uniform magnetic field can be seen in [13].

Also we suppose that a wavefunction $\psi(\mathbf{r},t)$ can be expressed in terms of the eigenfunctions $\psi_\lambda(\mathbf{r})$

$$\psi(\mathbf{r},t) = \sum_\lambda c_\lambda(t)\psi_\lambda(\mathbf{r}) \qquad (65)$$

with the coefficient $c_\lambda(t)$.

Then the kinetic operator $T(\mathbf{r},\hat{\mathbf{p}},t)$ is defined as

$$T(\mathbf{r},\hat{\mathbf{p}},t)\psi(\mathbf{r},t) = T(\mathbf{r},\hat{\mathbf{p}},t)\sum_\lambda c_\lambda(t)\psi_\lambda(\mathbf{r}) = \sum_\lambda \sqrt{\lambda(t) + m^2c^4}\, c_\lambda(t)\psi_\lambda(\mathbf{r}). \qquad (66)$$

Based on this definition, one can easily verify that $T(\mathbf{r},\hat{\mathbf{p}},t)$ is Hermitian, $T^+ = T$.

The Schrödinger equation is

$$i\hbar\frac{\partial}{\partial t}\psi(\mathbf{r},t) = H(\mathbf{r},\hat{\mathbf{p}},t)\psi(\mathbf{r},t), \qquad (67)$$

or

$$i\hbar\frac{\partial}{\partial t}\psi(\mathbf{r},t) = \sqrt{(\mathbf{p}c - q\mathbf{A}(\mathbf{r},t))^2 + m^2c^4}\,\psi(\mathbf{r},t) + q\Phi(\mathbf{r},t)\psi(\mathbf{r},t). \qquad (68)$$

Historically [11], due to the complication and difficulty of the square root operator, this equation has been abandoned, and the Klein-Gordon equation and the Dirac equation were developed to avoid the square root operator. However, we discovered that the square root operator was necessary to avoid the divergence in the perturbation method for the calculation on the relativistic correction of the hydrogen energy levels. What is more, the eigenenergy of the square root equation (54) has a special $\alpha^5$ term, which is 41% of the measured Lamb shift [3,4], while the energy levels from the Schrödinger equation, the Klein-Gordon equation, and the Dirac equation do not have an $\alpha^5$ term at all.

Readers can explicitly write out the probability continuity equation with or without sources according to the definitions in Sec. III.

## 3. Fractional Schrödinger equation

The fractional Schrödinger equation [5,6] is

$$i\hbar \frac{\partial}{\partial t}\psi(\mathbf{r},t) = H\psi(\mathbf{r},t), \tag{69}$$

with the fractional Hamiltonian operator

$$H = T_\alpha + V(\mathbf{r}). \tag{70}$$

We can prove that if the potential is even, $V(-\mathbf{r},t) = V(\mathbf{r},t)$, then the parity is conserved and that if the potential is central, $V(\mathbf{r},t) = V(r,t)$, then the angular momentum is conserved.

In the fractional quantum mechanics, the probability continuity equation can be written as

$$\frac{\partial}{\partial t}\rho + \nabla \cdot \mathbf{j} = 0 \tag{71}$$

with the probability density and the current density

$$\begin{aligned}\rho &= \psi^*\psi \\ \mathbf{j} &= \frac{i}{\hbar}\nabla^{-2}\nabla\left(\psi^*(T_\alpha\psi) - \psi(T_\alpha\psi)^*\right),\end{aligned} \tag{72}$$

or

$$\frac{\partial}{\partial t}\rho + \nabla \cdot \mathbf{j}_{\alpha 1} = I_{\alpha 1}, \tag{73}$$

where the current density and the source term are defined as

$$\mathbf{j}_{\alpha 1} = -iD_\alpha \hbar^{\alpha-1} \left( \psi^*(-\nabla^2)^{\alpha/2-1}\nabla\psi - \psi(-\nabla^2)^{\alpha/2-1}\nabla\psi^* \right), \tag{74}$$

$$I_{\alpha 1} = -iD_\alpha \hbar^{\alpha-1} \left( \nabla\psi^*(-\nabla^2)^{\alpha/2-1}\nabla\psi - \nabla\psi(-\nabla^2)^{\alpha/2-1}\nabla\psi^* \right). \tag{75}$$

In [6], it was a mistake that the source term $I_{\alpha 1}$ was missing [14].

## 4. The Schrödinger equation with a position-dependent mass

The deformed Schrödinger equation is [9,10]

$$i\hbar \frac{\partial}{\partial t}\psi(\mathbf{r},t) = H\psi(\mathbf{r},t), \tag{76}$$

where the deformed Hamiltonian operator

$$H = \mathbf{p}\cdot\frac{1}{2m(\mathbf{r})}\mathbf{p} + V(\mathbf{r}) \tag{77}$$

has a position-dependent mass $m(\mathbf{r})$.

We can prove that if the mass function and the potential are even, m(-**r**)=m(**r**) and $V(-\mathbf{r})=V(\mathbf{r})$, then the parity is conserved and that if the mass function and the potential is central, m(-**r**)=m(r) and $V(\mathbf{r})=V(r)$, then the angular momentum is conserved.

According to Sec III, the probability continuity equation without sources is

$$\frac{\partial}{\partial t}\rho + \nabla\cdot\mathbf{j} = 0 \tag{78}$$

where the current density is

$$\begin{aligned}
\mathbf{j} &= \frac{i}{\hbar}\nabla^{-2}\nabla\left(\psi^*(T\psi) - \psi(T\psi)^*\right) \\
&= \frac{i}{\hbar}\nabla^{-2}\nabla\left(\psi^*\mathbf{p}\cdot\frac{1}{2m(\mathbf{r})}\mathbf{p}\psi - \psi\mathbf{p}\cdot\frac{1}{2m(\mathbf{r})}\mathbf{p}\psi^*\right) \\
&= \nabla^{-2}\nabla\cdot\nabla\left(\psi^*\frac{1}{2m(\mathbf{r})}\mathbf{p}\psi - \psi\frac{1}{2m(\mathbf{r})}\mathbf{p}\psi^*\right) \\
&= \psi^*\frac{1}{2m(\mathbf{r})}\mathbf{p}\psi - \psi\frac{1}{2m(\mathbf{r})}\mathbf{p}\psi^*.
\end{aligned} \tag{79}$$

Readers can further study the probability continuity equation with sources.

## 5. The Dirac equation.

The Dirac equation is

$$i\hbar \frac{\partial}{\partial t} \psi(\mathbf{r},t) = H\psi(\mathbf{r},t), \tag{80}$$

where the Hamiltonian operator

$$H = c\vec{\alpha} \cdot \mathbf{p} + \beta mc^2 + V(r). \tag{81}$$

Please notice that now the wavefunction has 4 components.

**Formula 2.** If an eigenvalue and an eigenfunction of the Dirac's Hamiltonian operator are E and $\psi(\mathbf{r})$, i.e.

$$H\psi(\mathbf{r}) = E\psi(\mathbf{r}), \tag{82}$$

then we have

$$E = <V> + \sqrt{<\mathbf{p}^2 c^2> + m^2 c^4 + <V>^2 - <V^2>}. \tag{83}$$

**Proof.** The proof is similar to the proof of Formula 1, since from (81), again we have

$$H - V = c\vec{\alpha} \cdot \mathbf{p} + \beta mc^2 \tag{84}$$

$$(H-V)^2 = \mathbf{p}^2 c^2 + m^2 c^4 \tag{85}$$

$$H^2 - HV - VH + V^2 = \mathbf{p}^2 c^2 + m^2 c^4. \tag{86}$$

Readers can complete the proof easily. Again the average of the kinetic energy is

$$<c\vec{\alpha} \cdot \mathbf{p}> = \sqrt{<\mathbf{p}^2 c^2> + m^2 c^4 + <V>^2 - <V^2>}. \tag{87}$$

It is well know that the parity and the angular momentum are not conserved.

When there is the electromagnetic field, we have

$$H = c\vec{\alpha} \cdot (\mathbf{p} - q\mathbf{A}/c) + \beta mc^2 + V(r).$$

Based on the definition (33), the probability continuity equation without sources is

$$\frac{\partial}{\partial t} \rho + \nabla \cdot \mathbf{j} = 0 \tag{88}$$

where the probability and current density are defined as

$$\rho = \psi^+\psi^+$$
$$\mathbf{j} = \frac{i}{\hbar}\nabla^{-2}\nabla\left(\psi^+(c\vec{\alpha}\cdot(\mathbf{p}-q\mathbf{A}/c)+\beta mc^2)\psi - [(-c\vec{\alpha}\cdot(\mathbf{p}-q\mathbf{A}/c)+\beta mc^2)\psi]^+\psi\right)$$
$$= \frac{i}{\hbar}\nabla^{-2}\nabla\left(\psi^+c\vec{\alpha}\cdot\mathbf{p}\psi + \mathbf{p}\cdot\psi^+c\vec{\alpha}\psi\right) \qquad (89)$$
$$= c\nabla^{-2}\nabla\cdot\nabla\left(\psi^+\vec{\alpha}\psi\right)$$
$$= c\psi^+\vec{\alpha}\psi.$$

This is the popular form of the probability continuity equation. Based on the definition (35), we have

$$\frac{\partial}{\partial t}\rho_1 + \nabla\cdot\mathbf{j}_1 = I_1 \qquad (90)$$

with

$$\rho_1 = \psi^+\psi$$
$$\mathbf{j}_1 = 2c\psi^+\vec{\alpha}\psi \qquad (91)$$
$$I_1 = c\nabla\cdot\psi^+\vec{\alpha}\psi.$$

## V. An example of the non-Hermitian kinetic energy operator.

We will see that a non-Hermitian system is quite different from the well-known Hermitian system. Our discussion is based on an example, K-G equation without magnetic field.

The K-G equation for a spinless particle of mass m and charge q moving in an electromagnetic field is

$$\left(i\hbar\frac{\partial}{\partial t}-q\Phi\right)^2\psi(\mathbf{r},t) = (\mathbf{p}c-q\mathbf{A})^2\psi(\vec{r},t) + m^2c^4\,\psi(\vec{r},t)\,. \qquad (92)$$

In a static pure electrical field, $\mathbf{A}=0, q\Phi=V(\mathbf{r})$, we have

$$\left(i\hbar\frac{\partial}{\partial t}-V\right)^2\psi(\mathbf{r},t) = \mathbf{p}^2c^2\psi(\mathbf{r},t) + m^2c^4\,\psi(\mathbf{r},t). \qquad (93)$$

### 1. The Schrödinger equation form of the Klein-Gordon

We report that the K-G equation (93) can be expressed as a Schrödinger equation with a non-Hermitian kinetic energy operator,

$$i\hbar \frac{\partial}{\partial t} \psi(\mathbf{r},t) = H \psi(\mathbf{r},t),$$
$$H = T + V = \sqrt{\mathbf{p}^2 c^2 + m^2 c^4 + [V,H]} + V. \qquad (94)$$

If $[V,H]$ is zero, or so small that it can be neglected, this Hamiltonian and the relativistic Hamiltonian (51) is the same. Therefore, we need to pay more attention on the other cases.

Suppose the eigenequation

$$H\psi(\mathbf{r}) = E\psi(\mathbf{r}) \qquad (95)$$

has eigenfunctions $\psi_n(\mathbf{r})$ and eigenvalues $E_n$, where n is an index.

Then the general solution of the Schrödinger equation (94) is

$$\psi(\mathbf{r},t) = \sum_n c_n \psi_n(\mathbf{r}) \exp(-iE_n t), \qquad (96)$$

with the arbitrary coefficients $c_n$.

**Proposition 1.** The Schrödinger equation (94) implies the K-G equation (93).

**Proof.** The deduction from (94) to (93) is as follows

$$i\hbar \frac{\partial}{\partial t} \psi(\mathbf{r},t) = \sqrt{\mathbf{p}^2 c^2 + m^2 c^4 + [V,H]}\, \psi(\mathbf{r},t) + V \psi(\mathbf{r},t)$$

$$\left( i\hbar \frac{\partial}{\partial t} - V \right) \psi(\mathbf{r},t) = \sqrt{\mathbf{p}^2 c^2 + m^2 c^4 + [V,H]}\, \psi(\mathbf{r},t)$$

$$\sqrt{\mathbf{p}^2 c^2 + m^2 c^4 + [V,H]} \left( i\hbar \frac{\partial}{\partial t} - V \right) \psi(\mathbf{r},t) = \left( \mathbf{p}^2 c^2 + m^2 c^4 + [V,H] \right) \psi(\mathbf{r},t) \qquad (97)$$

$$\left( i\hbar \frac{\partial}{\partial t} - V \right) \sqrt{\mathbf{p}^2 c^2 + m^2 c^4 + [V,H]}\, \psi(\mathbf{r},t) = \left( \mathbf{p}^2 c^2 + m^2 c^4 \right) \psi(\mathbf{r},t)$$

$$\left( i\hbar \frac{\partial}{\partial t} - V \right)^2 \psi(\mathbf{r},t) = \left( \mathbf{p}^2 c^2 + m^2 c^4 \right) \psi(\mathbf{r},t).$$

This completes the proof.

## 2. The non-Hermiticity of the Hamiltonian

**Proposition 2.** The Hamiltonian operator is Hermitian if and only if $V(\mathbf{r}) = V_0$ a constant, i.e.

$$H^+ = H \iff V = V_0. \qquad (98)$$

**Proof**. The proof has two steps.

(1) On $\Leftarrow$.

Since $V(\mathbf{r}) = V_0$, we have $[V, H] = 0$, and

$$H = T + V = \sqrt{\mathbf{p}^2 c^2 + m^2 c^4} + V_0. \tag{99}$$

This is a Hermitian operator.

(2) On $\Rightarrow$.

Now H is a Hermitian operator, i.e., $H^+ = H$.

From (94) we have

$$H - V = \sqrt{\mathbf{p}^2 c^2 + m^2 c^4 + [V, H]} \tag{100}$$

$$(H - V)^2 = \mathbf{p}^2 c^2 + m^2 c^4 + [V, H] \tag{101}$$

$$H^2 - 2VH + V^2 = \mathbf{p}^2 c^2 + m^2 c^4. \tag{102}$$

Taking the Hermitian adjoint of the two sides, we have

$$H^{2+} - 2H^+ V + V^2 = \mathbf{p}^2 c^2 + m^2 c^4 \tag{103}$$

$$H^2 - 2HV + V^2 = \mathbf{p}^2 c^2 + m^2 c^4. \tag{104}$$

Taking (102) - (104), we have

$$HV = VH, \quad [V, H] = 0. \tag{105}$$

Therefore, we have

$$H = \sqrt{\mathbf{p}^2 c^2 + m^2 c^4} + V, \tag{106}$$

and

$$[H, V] = [\sqrt{\mathbf{p}^2 c^2 + m^2 c^4}, V] = 0. \tag{107}$$

Further, we have

$$\begin{aligned}[\mathbf{p}^2 c^2 + m^2 c^4, V] \\ = \sqrt{\mathbf{p}^2 c^2 + m^2 c^4}\left[\sqrt{\mathbf{p}^2 c^2 + m^2 c^4}, V\right] + \left[\sqrt{\mathbf{p}^2 c^2 + m^2 c^4}, V\right]\sqrt{\mathbf{p}^2 c^2 + m^2 c^4} = 0,\end{aligned} \tag{108}$$

$$[\mathbf{p}^2, V] = 0. \tag{109}$$

Therefore, the potential energy is a constant

$$V(\mathbf{r}) = V_0. \tag{110}$$

This completes the proof.

In fact, based on Eq. (103), we have

$$\begin{aligned} H^+ &= T^+ + V = \sqrt{\mathbf{p}^2 c^2 + m^2 c^4 + [H^+, V]} + V \\ &= \sqrt{\mathbf{p}^2 c^2 + m^2 c^4 + [V, H]^+} + V. \end{aligned} \tag{111}$$

### 3. The average formula

**Formula 3.** If an eigenvalue and an eigenfunction of the Hamiltonian operator are E and $\psi(\mathbf{r})$, i.e.

$$H\psi(\mathbf{r}) = E\psi(\mathbf{r}), \tag{112}$$

then we have

$$E = <V> + \sqrt{<\mathbf{p}^2 c^2> + m^2 c^4 + <V>^2 - <V^2>}. \tag{113}$$

**Proof.**

Taking the integral $\int_{R^3} \psi^*(\mathbf{r}) \cdots \psi(\mathbf{r}) d^3\mathbf{r}$ of the two sides of Eq. (102), we have

$$\int_{R^3} \psi^*(\mathbf{r})(H^2 - 2VH + V^2)\psi(\mathbf{r}) d^3\mathbf{r} = \int_{R^3} \psi^*(\mathbf{r})(\mathbf{p}^2 c^2 + m^2 c^4)\psi(\mathbf{r}) d^3\mathbf{r} \tag{114}$$

Using the Dirac's notations, we have

$$<H^2> - 2<VH> + <V^2> = <\mathbf{p}^2 c^2> + m^2 c^4$$

$$E^2 - 2E<V> + <V^2> = <\mathbf{p}^2 c^2> + m^2 c^4$$

$$(E - <V>)^2 = <\mathbf{p}^2 c^2> + <V>^2 - <V^2> + m^2 c^4 \tag{115}$$

Usually the right-hand side of Eq. (115) is greater than zero for a particle with a low speed. Taking the positive solution, we have the energy formula

$$E = <V> + \sqrt{<\mathbf{p}^2 c^2> + <V>^2 - <V^2> + m^2 c^4}. \tag{116}$$

If the right-hand side of Eq. (115) is smaller than zero, we see that the energy E will become a complex number.

Again, the value of the kinetic energy operator is

$$<\sqrt{\mathbf{p}^2c^2 + m^2c^4 + [V,H]}> = \sqrt{<\mathbf{p}^2c^2> + m^2c^4 + <V>^2 - <V^2>}. \tag{117}$$

Notice that Formulas 1, 2, and 3 have the same form and can be considered as one formula.

## 4. The non-orthogonality of the eigenfunctions

**Proposition 3.** Two eigenfunctions $\psi_1(\mathbf{r}), \psi_2(\mathbf{r})$, whose eigenenergies $E_1$, $E_2$ satisfy $E_2^{*2} \neq E_1^2$, are orthogonal if and only if

$$\int_{R^3} \psi_1^*(\mathbf{r}) V \psi_2(\mathbf{r}) d^3\mathbf{r} = 0. \tag{118}$$

**Proof.** From Eq. (102) we have

$$H^2\psi_2 - 2VH\psi_2 + V^2\psi_2 = \mathbf{p}^2c^2\psi_2 + m^2c^4\psi_2$$
$$E_2^2\psi_2 - 2E_2V\psi_2 + V^2\psi_2 = \mathbf{p}^2c^2\psi_2 + m^2c^4\psi_2$$

$$E_2^2\psi_1^*\psi_2 - 2E_2\psi_1^*V\psi_2 + \psi_1^*V^2\psi_2 = \psi_1^*\mathbf{p}^2c^2\psi_2 + m^2c^4\psi_1^*\psi_2. \tag{119}$$

Similarly, we have

$$E_1^2\psi_2^*\psi_1 - 2E_1\psi_2^*V\psi_1 + \psi_2^*V^2\psi_1 = \psi_2^*\mathbf{p}^2c^2\psi_1 + m^2c^4\psi_2^*\psi_1. \tag{120}$$

Taking the complex conjugate, we have

$$E_1^{*2}\psi_1^*\psi_2 - 2E_1^*\psi_1^*V\psi_2 + \psi_1^*V^2\psi_2 = \psi_2\mathbf{p}^2c^2\psi_1^* + m^2c^4\psi_1^*\psi_2. \tag{121}$$

Taking Eq. (119) - Eq. (121), we have

$$(E_2^{*2} - E_1^2)\psi_1^*\psi_2 - 2(E_2^* - E_1^*)\psi_1^*V\psi_2 = \psi_1^*\mathbf{p}^2c^2\psi_2 - \psi_2\mathbf{p}^2c^2\psi_1^*. \tag{122}$$

Taking the integral $\int_{R^3} \cdots d^3\mathbf{r}$, we have

$$(E_2^{*2} - E_1^2)\int_{R^3}\psi_1^*\psi_2 d^3\mathbf{r} - 2(E_2^* - E_1^*)\int_{R^3}\psi_1^*V\psi d^3\mathbf{r} = 0 \tag{123}$$

$$(E_2^* + E_1)\int_{R^3}\psi_1^*\psi_2 d^3\mathbf{r} - 2\int_{R^3}\psi_1^*V\psi_2 d^3\mathbf{r} = 0 \tag{124}$$

$$\int_{R^3} \psi_1^* \psi_2 d^3\mathbf{r} = \frac{2}{E_2^* + E_1} \int_{R^3} \psi_1^* V \psi_2 d^3\mathbf{r} . \tag{125}$$

It is easy to see that $\psi_1(\mathbf{r}), \psi_2(\mathbf{r})$ are orthogonal if and only if

$$\int_{R^3} \psi_1^* V \psi_2 d^3\mathbf{r} = 0 . \tag{126}$$

This completes the proof.

## 5. The Non-conservation of physical quantities

Based on Theorem 1, the quantity

$$\int_{R^3} \psi^*(\mathbf{r},t) H \psi(\mathbf{r},t) d^3\mathbf{r}$$

is not conserved generally. Since H is not Hermitian (except the trivial case $V(\mathbf{r}) = V_0$), there must exist some wavefunctions $\psi(\mathbf{r})$ such that this integral is a complex number [11]. For these two reasons we know that H is not an operator associated to the total energy.

Strictly speaking, the physics meanings of the kinetic energy operator and the Hamiltonian operator are not clear. It is not suitable to call the eigenvalues of H (e.g. Eq. (10) for the Coulomb potential) the energy levels; they are just a parameters with the dimension of the energy appearing in the solution (96).

Based on Theorem 3 the integral $\int_{R^3} \psi^*(\mathbf{r},t) \psi(\mathbf{r},t) d^3\mathbf{r}$ is not necessarily conserved. Still, readers can study the probability continuity equation according to the formulas in Section III.

We can say that if the potential is even, $V(-\mathbf{r}) = V(\mathbf{r})$, then

$$[H, \mathbf{P}] = 0, \tag{127}$$

since

$$\begin{aligned} H &= H(\mathbf{p}^2, V(\mathbf{r})), \\ [\mathbf{p}^2, P] &= 0, \quad [V(\mathbf{r}), P] = 0. \end{aligned} \tag{128}$$

If the potential is central, $V(\mathbf{r}) = V(r)$, then

$$[H, \mathbf{L}] = 0 \tag{129}$$

since

$$H = H(\mathbf{p}^2, V(r)),$$
$$[\mathbf{p}^2, \mathbf{L}] = 0, \quad [V(r), \mathbf{L}] = 0. \tag{130}$$

From (129), we know that

$$[H, \mathbf{L}^2] = 0, \ [H, L_z] = 0, \ [\mathbf{L}^2, L_z] = 0. \tag{131}$$

Thus in the central potential, one can find the common eigenfunctions of the sets $H, \mathbf{L}^2, L_z$. For example the solution to the K-G equation with a Coulomb potential (9) can be seen in [1,2].

However, the parity and the angular momentum are not conserved anymore.

We see that the quantum mechanics based on K-G equation is quite different from the standard quantum mechanics. For these reasons, we think it would be better to use the relativistic Schrödinger equation (52) than the K-G equation (93) to describe the spinless particle, e.g. pionic hydrogen atom (a pion moving in the Coulomb field of a proton).

## 6. Another non-Hermitian Hamiltonian related to the K-G equation

In addition, we mention another non-Hermitian kinetic energy operator, which also relates to the K-G equation.

**Proposition 4**. The Schrödinger equation

$$i\hbar \frac{\partial}{\partial t} \psi(\mathbf{r}, t) = H' \psi(\mathbf{r}, t)$$
$$H' = T' + V' = -\sqrt{\mathbf{p}^2 c^2 + m^2 c^4 + [V', H']} + V' \tag{132}$$

implies the K-G equation (93) as well.

**Proof.** The proof is similar to that of Proposition **1.**

**Proposition 5**. If $V' = -V$, then $H' = -H$.

Proof. If $V' = -V$, according to (132), we have

$$H' = -\sqrt{\mathbf{p}^2 c^2 + m^2 c^4 + [-V, H']} - V \tag{133}$$

$$-H' = \sqrt{\mathbf{p}^2 c^2 + m^2 c^4 + [V, -H']} + V \tag{134}$$

Comparing (134) with the definition of H in (94), we have $H = -H'$. This completes the proof.

**Exmaple**. The two operators

$$H = \sqrt{\mathbf{p}^2c^2 + m^2c^4 + \left[-e^2/r, H\right]} - e^2/r \tag{135}$$

$$H' = -\sqrt{\mathbf{p}^2c^2 + m^2c^4 + \left[e^2/r, H'\right]} + e^2/r \tag{136}$$

have the same eigenfunctions and opposite eigenenergies.

## VI. Conclusion

The Schrödinger equation with a general kinetic energy operator is introduced. The conservation law is proved and the probability continuity equation is deducted in a general sense. Examples with a Hermitian kinetic energy operator include the standard Schrödinger equation, the relativistic Schrödinger equation, the fractional Schrödinger equation, the deformed Schrödinger equation, and the Dirac equation. We reveal that the Klein-Gordon equation has a hidden non-Hermitian kinetic energy operator. The probability continuity equation with sources indicates that there exists a different way of the probability transportation, which is probability teleportation. An average formula is deducted from the relativistic Schrödinger equation, the Dirac equation, and the K-G equation.

### Acknowledgement

The research on the relativistic Schrödinger equation was supported by Gansu Industry University (currently Lanzhou University of Technology) during 1989-1991, with a project title "On the solvability of the square root equation in the relativistic quantum mechanics".

The author thanks Reviewer B for his correction on Sec III.2 and reminding me of some related monographs.

**Appendix**. A discussion on the relativistic Schrödinger equation (52).

Reviewer A:

*The relativistic Schrödinger equation (52) is not relativistically invariant at all, and was ruled out in the early days of relativistic quantum mechanics, see [15]. Why do you pay attention to such a wrong and useless equation?*

Response.

First of all, this equation does not have an official name [11,15]. The name relativistic Schrödinger equation here means the Schrödinger equation with a relativistic kinetic energy operator. I just noticed that this name is used in [16] as well.

Based on my own research, there are two new reasons to revive this equation. (1) This equation appears naturally when calculating the relativistic kinetic correction of the energy levels of the H atom by the perturbation method [3, 4]. (2) This equation is an approximate realization of the well-known fractional Schrödinger equation [7,8].

Additionally, I want to add the following two points.

1. Relativistic covariance and correctness

Indeed this equation is not relativistically invariant, but the standard Schrödinger equation is not either. Therefore a non-covariant equation can be valuable if the equation is practically useful, since a non-covariant equation can become an approximation of a covariant equation somehow in the future, as the Schrödinger equation becomes an approximation of the Dirac equation and the Klein-Gordon equation.

2. Experimental criterion.

The final criterion to judge the value of an equation is whether the equation can predict new experimental results.

(1) The energy formula for the Hydrogen atom based on this equation is even better than that from the Dirac equation. In addition, I wish that the experimental physicists can soon judge which equation generates the correct energy level formula for the pionic hydrogen atom [1,2], the relativistic Schrödinger equation or the K-G equation [3]. (2) The relativistic Schrödinger equation indicates the possibility of probability teleportation, quite different from the Schrödinger equation, the K-G equation, and the Dirac equation. We are designing related experiments to observe this new relativistic quantum mechanics phenomenon.

If the two experimental results did not support the relativistic Schrödinger equation, I would admit that this equation is wrong both theoretically and experimentally.